# Opinion Dynamics of Online Social Network Users: A Micro-Level Analysis

**(This is an original manuscript of an article published by Taylor & Francis in The Journal of Mathematical Sociology on August 4, 2021, available online: [http://www.tandfonline.com/10.1080/0022250X.2021.1956917](http://www.tandfonline.com/10.1080/0022250X.2021.1956917))**

Ivan V. Kozitsin[1, 2]

*[1]Laboratory of Complex Networks, V. A. Trapeznikov Institute of Control Sciences of Russian Academy of Sciences, 65 Profsoyuznaya street, Moscow, 117997, Russian Federation*

*[2]Department of Higher Mathematics, Moscow Institute of Physics and Technology, 9 Institutskiy per., Dolgoprudny, Moscow Region, 141701, Russian Federation*

**Email**: kozitsin.ivan@mail.ru

**Funding**: this work was supported by the Russian Foundation for Basic Research, project no. 18-29-22042.

In this paper, we present an empirical study of the opinion dynamics of a large-scale sample of online social network users. We estimate users' opinions as continuous scalars based on their subscriptions to information sources and analyze how friendship connections affect the dynamics of these estimations. Distinguishing between positive (toward friends' opinions) and negative (away from friends' opinions) opinion shifts, we find that the existence and magnitude of opinion shifts are positively related (largely through a linear form or an inverted U-shaped form) to the degree of divergence in opinions between user and their friends. Additionally, we moderate the balance between positive and negative shifts using opinion divergence: if the opinions of the focal user and their friends are too similar or dissimilar, there is a relatively low chance of a positive shift.

## Introduction

Individuals change their views and behavior over time and a remarkable contribution to this process is provided by social influence, a force that affects people through their peers and other sources of information. Today, online social networks (OSNs) have become a crucial channel of information diffusion. Users of these platforms can broadcast their views effectively, without the restrictions inherent in older forms of communication. They have given rise to large-scale systems of influence processes that unfold in an online environment. Investigation of such social systems is a challenging task that can, nonetheless, bring new insight into our knowledge of how individuals form their opinions (Bond et al., 2012; Ravazzi et al., 2020). By observing the opinion dynamics of users at the micro-scale, scholars can identify key micro-assumptions regarding the process of opinion formation and integrate this knowledge into existing opinion formation models (Abid et al., 2018; Castellano et al., 2009; Flache et al., 2017; Friedkin et al., 2016; Mäs, 2019; Proskurnikov et al., 2017; Proskurnikov & Tempo, 2017, 2018).

In this paper, we analyze longitudinal data on the opinion dynamics of a large-scale (1.6 M) sample of OSN users. Handling both users' opinions (based on the information sources to which they are subscribed) and the structure of users' friendship ties, we test competing assumptions on individuals' opinion shifts against real data.

The majority of existing empirical studies suggest that the formation of continuous opinions (presented as continuous quantities) largely follow the linear positive mechanism whereby individuals tend to modify their opinions toward those of their friends with a magnitude proportional to the pre-existing difference in opinions (Friedkin et al., 2021; Takács et al., 2016). However, some papers suggest a moderated positive influence: there is a maximum rate of influence if the difference is not too small and not too high; otherwise, it decreases (Moussaïd et al., 2013). Studies that reported a negative influence (opinion shifts away from the opinions of peers) either have methodological concerns (Knippenberg et al., 1990; Mazen & Leventhal, 1972) or are based on natural experiments in which researchers are unable to control for all possible confounding factors (Liu & Srivastava, 2015).

Our contributions are as follows. We distinguish between positive (toward friends' opinions) and negative (away from friends' opinions) opinion shifts. We find that the properties of positive shifts are largely in line with the predictions of existing theoretical models of opinion formation. Precisely, the existence and magnitude of positive opinion shifts are positively related (largely through a linear form or an inverted U-shaped form) to the degree of divergence in opinions between user and their friends. Interestingly, we find that both the chance and magnitude of negative shifts increase alongside the opinion divergence between the focal user and their friends (largely through, again, a linear or an inverted U-shaped form). The balance between positive and negative shifts is also moderated by opinion divergence: if the opinions of the focal user and their friends are too similar or dissimilar, there is a relatively low chance of a positive shift. These results challenge existing concerns about the presence of a negative influence; they constitute a clue for the context in which researchers should search for such an influence. However, we also demonstrate how to explain our empirical results without referring to the notions of positive and negative influence. Concurrently, we investigate how the process of radicalization is connected to the ideological heterogeneity of users' neighborhoods. We argue that individuals are less likely to radicalize if they are exposed to moderate positions; in contrast, radical views—regardless of their bias—induce a relatively high degree of radicalization.

The remainder of this paper is organized as follows. Section 1 reviews the study's foundational theory. In Section 2, we discuss the data at hand. Sections 3–5 present our main results. In Section 6, we discuss our results. Section 7 offers a conclusion and suggests potential avenues for future research. The Appendix includes supplemental information.

## 1. Theory

**Social influence mechanisms.** The literature on social influence models (those in which opinions are conceptualized as continuous quantities) emphasizes three main influence mechanisms (Flache et al., 2017): (1) positive (assimilative) influence; (2) positive influence coupled with the bounded confidence assumption; and (3) negative (repulsive) influence (see Figure 1). Here, we briefly discuss these three mechanisms.

[Figure 1 is near here]

In the case of positive influence, agent $i$ aligns their opinion more closely with that of influence source $j$. If opinions are represented on a one-dimensional axis (the only situation that we consider), agent $i$'s new opinion lands somewhere between influence source $j$'s opinion and agent $i$'s initial opinion (DeGroot, 1974) (see Figure 1, panel a). This opinion-formation mechanism has gained remarkable attention from the scientific community, in which it is sometimes referred to as the *weighted averaging mechanism* (Proskurnikov & Tempo, 2017). Another possibility that researchers may encounter while analyzing real data—despite rarely garnering attention in the literature—is the leapfrogging of a newly formed opinion past that of the influence source (Friedkin et al., 2021) (Figure 1, panel b). Of course, such "skips" could simply stem from measurement errors or external sources of influence that were not considered by researchers.

However, positive influence (without skips) in a typical case leads to an inevitable consensus in a connected social system; this result cannot explain persistent disagreement or opinion polarization in both small groups and large populations (Abelson, 1964). For this reason, the *bounded confidence* mechanism has been introduced, whereby agent $i$ is influenced only if the opinion of influence source $j$ is sufficiently similar to that of agent $i$ (Deffuant et al., 2000; Hegselmann & Krause, 2002); influence source $j$'s opinion must lie in agent $i$'s *interval of confidence* (Figure 1, panel c). Simulation experiments have revealed that, if the intervals of confidence are small enough in a bounded confidence setting, a social system may reach an equilibrium state in which several opinion clusters coexist. Criticizing for its severity—due to the fact that it excludes interactions between agents with highly dissimilar opinions (Mäs et al., 2010)—researchers have advanced the bounded confidence mechanism by adding the possibility of opinion exchange, albeit a relatively minor one, between agents with polar positions: the *relaxed bounded confidence* assumption. However, additional ingredients should be added to generate robust disagreement within such a setting, such as "striving for uniqueness" (Mäs et al., 2010) or some level of randomness (that may stand for external influences) (Kurahashi-Nakamura et al., 2016).

It is important to note that there are other mechanisms that can explain persistent disagreement in groups. For example, incorporating *stubborn agents*, which are not sensitive to influence and do not change their positions, into the model can prevent a consensus state (Proskurnikov & Tempo, 2018). A conceptually different (but mathematically similar) approach is to consider prejudices, which exert a permanent influence on individuals and, in contrast to opinions, do not change over time (Friedkin, 2006; Friedkin & Johnsen, 1990).

However, all of these models still fail to explain polarization—the growth in opinion divergence over time, whereby individuals' opinions may even leave their initial interval (Dandekar et al., 2013). One potential solution here is the introduction of a negative influence whereby individuals with highly dissimilar opinions only intensify in divergence following an interaction (Altafini, 2013; Flache & Mäs, 2008; Macy et al., 2003) (Figure 1, panel d). In contrast to positive and bounded confidence influence mechanisms, which have substantial empirical support (Asch & Guetzkow, 1951; Nickerson, 1998; Sherif & Hovland, 1961; Takács et al., 2016), the assumption of a negative influence suffers from a lack of definitive empirical studies in its favor (Mäs & Flache, 2013; Takács et al., 2016).

Other examples of models explaining polarization without the assumption of a negative influence can be found in Banisch and Olbrich (2019), Dandekar et al. (2013), and Mäs and Flache (2013). They are built on the idea of biased assimilation—the tendency of individuals to seek information that aligns with their views, avoid cross-cutting content, and interpret messages in a comfortable manner (Dandekar et al., 2013;

Nickerson, 1998)—or on the introduction of more subtle forms of communication, such as specific arguments.

**Mathematical formalization.** Almost all of the abovementioned models can be captured by the following minimal opinion-formation rule:

$$x_i(t+1) = x_i(t) + l\Big(x_j(t) - x_i(t)\Big). \qquad (1)$$

In equation (1), agent $i$'s opinion at time $t+1$ (assumed to be a scalar from a continuous one-dimensional space) is formed as the sum of their past opinion and the term standing for the influence from source $j$ (representing the overall influence on $i$). The time scale in equation (1) is typically unspecified, and one step may stand for various timespans (e.g., day, week, month). The quantity $l$ describes how agent $i$ responds to the social influence, covering a wide range of micro-assumptions about influence processes. For simplicity, we do not bind opinions to a predefined interval. On the one hand, the opinion-formation rule in equation (1) can be understood as a formalization of a pairwise (one-to-one) interaction between agent $i$ and influence source $j$. On the other hand, the quantity $x_j(t)$ may stand for the overall influence of peers on agent $i$ (many-to-one interaction). One could further clarify $x_j(t)$ by noting, for example, that $x_j(t)$ is a convex combination of the opinions held by $i$'s peers.

From equation (1), it follows that $l$ describes how the shift in agent $i$'s opinion is connected with the opinion divergence between $i$ and $j$:

$$l = \frac{x_i(t+1) - x_i(t)}{x_j(t) - x_i(t)}$$

If $l$ is a positive constant (see Figure 2, which we borrowed from Takács et al. (2016)), we get a *positive linear influence* dependency and the corresponding assumption of positive linear influence. Note that, to avoid skips, we should require $l \leq 1$. It is worthwhile to note that the social judgment theory posits that the power of influence should decline when the opinion divergence between $i$ and $j$ is too small or too large (Sherif & Hovland, 1961). Additionally, the relaxed bounded confidence assumption prohibits influence when the distance between opinions is too large. As such, one may expect a *moderated positive influence* relation whereby the curve illustrating agent $i$'s opinion shift—which depends on the distance between $i$'s and $j$'s opinions—has an inverted U-shaped (negative quadratic) form (Takács et al., 2016). Assuming $l < 0$, we see a negative influence when $i$'s opinion shifts away from that of $j$. As with a positive influence, there may be *negative linear influence* and *moderated negative influence* dependencies that can be obtained from their positive counterparts through reflection on a horizontal axis (for brevity, we do not plot them). Note that the negative linear influence dependency is largely affected by the geometrical properties of the opinions space. For example, Takács et al. (2016) consider the case when opinions are restricted to lie in the interval $[0,100]$. As a result, they have a different formalization of the negative linear influence assumption. In Subsection 5.2, we return to this problem. However, negative influence is rarely modeled without positive influence, so one potential solution here is to combine the moderated positive and moderated negative influence components; when absolute opinion divergence becomes too substantial, the positive influence gives way to a negative influence that first grows in power but then disappears as individuals become so distinct in their views that there is no chance for interaction.

[Figure 2 is near here]

One question of great importance is as follows: under what conditions do individuals modify their opinions? Formal models of social influence typically assume that agent $i$ modifies their opinion (is active) so long as there is at least one influence source $j$ in their neighborhood with a different opinion. Both stubbornness and the bounded confidence assumption are potential solutions to control for the possibility that individuals hold their initial position despite receiving challenging information. In principle, one can simply extend the mentioned ideas by, for example, taking the absolute values of the corresponding curves from Figure 2 and, from them, deriving the similar dependencies for agents' activation probabilities. Under the positive linear influence assumption, the probability of an agent's activation should then grow alongside the level of discrepancy between them and their social environment. Analogously, under the moderated positive influence assumption, the agent is likely to change their opinion if the quantity $|x_i(t) - x_j(t)|$ is not too small and not too large. For the combination of the moderated positive and moderated negative influences, the dependency is richer and should have two global maxima that correspond to different influence types. Another potential way to control for agents' activation patterns is to consider the fact that individuals who are confident in their positions are hardly influenced (Moussaïd et al., 2013). The same holds for those who are highly appraised by their peers (Friedkin, 2011).

**Testing against real data.** Of course, it is important for formal models to be able to successfully describe real social processes (Flache et al., 2017; Mäs, 2019). Existing empirical studies (see Table 1) largely support the positive linear influence assumption; some of them also support the moderated positive influence assumption (Moussaïd et al., 2013), and some even support the existence of a negative influence (Knippenberg et al., 1990; Liu & Srivastava, 2015; Mazen & Leventhal, 1972). However, it is important to note that the studies that suggest a negative influence either have methodological concerns (Knippenberg et al., 1990; Mazen & Leventhal, 1972) or are based on natural experiments, which are more prone to improper conclusions than experiments in laboratory settings (Liu & Srivastava, 2015). However, it would be difficult to reproduce meaningful negative relationships, which constitute an essential source of repulsive dynamics, in a laboratory setting (Takács et al., 2016).

[Table 1 is near here]

This paper investigates the longitudinal data on the opinion dynamics of OSN users. This data has previously been analyzed by Kozitsin (2020), who concentrated on how users—endowed with estimated opinions from the interval [0,1]—choose the direction in which they will move in the opinion space (left or right). He finds that, after controlling for user opinion, the probability of shifting left varies with the average opinion of the user's friends, $x_{-i}$; the target function itself has a near-sinusoidal, wave-shaped form. If source opinion $x_{-i}$ is located more to the left than user opinion $x_i$ ($x_{-i}$ is less than $x_i$), the probability is greater than as for $x_{-i} > x_i$. In this paper, we explore the remaining important questions pertaining to opinion-formation processes: when—and to what degree—do individuals change their opinions?

## 2. Data Description

The data used in this paper initially appeared in Kozitsin et al. (2019); it was gathered from VKontakte (VK), the most popular OSN in Russia.

**Short overview of VK.** In terms of functionality, VK is very similar to Facebook. VK users can form follow-type (directed) connections with other users. After user $i$ establishes a follow-type connection with user $j$, user $j$ receives a notification indicating that user $i$ requests friendship. If user $j$ accepts, the two users become friends. Otherwise, user $i$ simply remains a follower of user $j$. Users with more than 1,000 followers have a special status on VK (henceforth referred to as *bloggers*): information on followed bloggers is displayed on user $i$'s account page in a special section. Users can also follow non-user accounts, including those for *groups*, *public pages*, and *events*. The main difference between public pages (that are also visible on user $i$'s account page in the same section as bloggers) and other non-user account types is that, while users may freely follow and unfollow public pages without restrictions, group-type accounts are able to limit access to their content. Bloggers and public pages are considered to be the main information disseminators on VK. It is worthwhile to note that the majority of media accounts on VK are public pages. VK Users receive new information from their news feeds, where content produced by the accounts to which they are subscribed is listed.

The data consists of three opinion snapshots from a sample of $N = 1,660,927$ VK users. This sample was obtained through random selection among active users (at least one platform interaction per month) who are Russians of at least 18 years of age with open privacy settings. Additionally, they needed to follow at least ten and at most 200 accounts of bloggers and public pages (hereafter – information sources). We explain the rationale behind this filter below. Users with no friends were excluded from the sample. The collected opinion snapshots represent users' attitudes toward President Putin in February ($t_1$), July ($t_2$), and December 2018 ($t_3$).

Using a supervised machine learning methodology introduced by Kozitsin et al. (2020), we estimated users' opinions as non-negative scalars, $x(t) \in [0,1]$. The quantity $x_i(t)$ represents user $i$'s opinion and is, generally speaking, a projection from the set of followed information sources to the interval $[0,1]$. The main idea of the estimation algorithm is that users' subscriptions should reflect their political views (Barberá, 2014; Frey, 1986; Tang & Chorus, 2019). If $x_i(t) = 0$, then user $i$ is a strong oppositionist. In contrast, $x_i(t) = 1$ means that this individual is a staunch supporter of President Putin. Meanwhile, users with $x_i(t) = 0.5$ are moderate individuals in terms of the opinions' assessing strategy. More precisely, such users are those who follow no politically relevant information sources or those who follow sources pertaining to polar ideological sides simultaneously. The argument $t$ in $x_i(t)$ reflects the moment in time at which the data used to produce the estimation was gathered. In our case, the time step $t \to t+1$ from equation (1) corresponds to a half-year step from $t_k \to t_{k+1}$. In the analysis that follows, we drop the time argument if it is already made clear by the context.

As pointed out by Kozitsin et al. (2020), estimations of users' opinions are most accurate when the number of their subscriptions on information sources is between ten and 200. To ensure sustainable opinion estimation, users who do not meet this requirement were eliminated from the sample.

The dataset also includes an adjacency matrix $A = [a_{ij}] \in \{0,1\}^{N \times N}$ that represents friendship connections between users from the sample. These connections were gathered in July 2018. The sets of information accounts and sample users have no intersections, facilitating the independence of users' opinion estimations from interconnections. Our analysis of the friendship network reveals that it includes a giant connected component of 1,648,829 nodes (99.3% of all nodes) and many (5535) tiny connected components (5,535) formed primarily by two or three nodes. In what follows, we focus on users from the giant connected component.

All of the data, codes, and other support information used in this study can be found at https://doi.org/10.7910/DVN/H3ZBHR.

## 3. Ideological Groups

We group individuals into five categories based on their opinions. For simplicity, we refer to individuals as "strong liberals" (SLs: $x_i \in [0,0.2)$), "liberals" (Ls: $x_i \in [0.2,0.4)$), "moderates" (Ms: $x_i \in [0.4,0.6)$), "conservatives" (Cs: $x_i \in [0.6,0.8)$), and "strong conservatives" (SCs: $x_i \in [0.8,1]$). Moving forward, individuals with opinions near the edges of the opinion space are referred to as those with *strong* or *radical* opinions. We will say that the opinions of SLs are *stronger* or *more radical* than those of Ls; Ms are the individuals with the *weakest* or *least radical* opinions. If $x_i < x_j$, then user $i$ is said to be more liberal than user $j$. We refer to users with opinions less than 0.5 as for those with *liberal bias*. Conversely, users having opinions greater than 0.5 are individuals with *conservative bias*. While we recognize that such names of groups are not correct from a political science theory standpoint, the analogies remain pertinent. We report that there are no users with opinions strictly equal to 0.5. The populations of each group (at time $t_2$) are presented in Table 2. We find that, on average, more liberal users have more friends (Pearson correlation coefficient equal to -0.16, p-value approximately equal to zero). To explain this result, we put forward the following hypothesis: liberal individuals are, on average, younger; thus, they should be more active on VK. We perform an additional experiment that partially confirms our hypothesis (see the Appendix for more details).

[Table 2 is near here]

## 4. Homophily Structure

Homophily is a well-documented phenomenon of social networks. It can be formulated as follows: if one takes a snapshot of a social network, they can observe that characteristics of connected individuals are more similar than one could expect in the case of randomly created ties (McPherson et al., 2001; M. Newman, 2018). Regarding opinions and behaviors, there are two principal explanations of such phenomena (Holme & Newman, 2006): selectivity (the tendency of individuals to form connections with those having similar traits, including opinions or behaviors) and social influence. For the analysis of opinion dynamics, it is essential to explore whether the system at hand is homophilic and to what extent as it (1) may be a result of the social influence processes (and, thus, may be used to make relevant hypotheses; see discussion at the end of this section) and (2) may affect how the influence processes unfold.

Kozitsin (2020) reports that the network under consideration is homophilic with the assortativity coefficient of approximately 0.14. In this study, we perform a comprehensive analysis of homophily patterns. For brevity, we concentrate on time moment $t_2$, at which the data on friendship connections was gathered. Using this data, we calculate the ideological composition of each user's neighborhood (represented as the fraction of their peers falling in each of five groups) and average these values across all ideological categories. We then compare our findings against the null model, in which connections between individuals appear at random. Within the null model, the list of friends of a randomly chosen user should consist of eight percent SLs, 19 percent Ls, 53 percent Ms, 16 percent Cs, and four percent SCs. If the system is homophilic, one can expect an enormous concentration of ties on and near the main diagonal.

Our analysis (see Table 3) reveals that users' tendency to have homophilic ties is strongly connected to their degree of radicalism: individuals with strong positions tend to have more ties with those holding similar views (than predicted by the null model). For moderate individuals, we observe little to no homophily. After controlling for the total number of friends, we find that for individuals having more friends, the connections map is significantly biased toward liberals (see Tables B1–B3 in the Appendix). We explain this result as follows: individuals with more friends tend to be younger and, thus, they should have younger friends (since social networks are assortative with respect to age) that, in turn, have more chances to be liberal.

[Table 3 is near here]

In general, our results indicate that the level of homophily varies across ideological groups. As such, we must carefully disentangle individuals with different positions while analyzing patterns of social influence. The following question naturally arises: "If the system at hand stems from selectivity or social influence (or a combination of both), what assumptions should we incorporate into models of selectivity or social influence (or both) to get results similar to those from Table 3?" In Section 6, we return to this problem.

## 5. Analysis of Opinion Shifts

In this section, we analyze users' opinion shifts from the perspective of the baseline micro-assumptions detailed in Section 1. We base our analysis on three principal aspects of opinion-formation processes:

**Q1**. Chance of opinion shift

**Q2**. Direction of opinion shift

**Q3**. Magnitude of opinion shift

**Q2** was covered extensively by Kozitsin (2020). As a result, instead of **Q2**, we address the specific issue of opinion radicalization (**Q2'**), which is of considerable interest in the literature: under what conditions are individuals more eager to become more radical in their views?

### 5.1. Preliminaries

Before starting the analysis, we must define some metrics. We use the average opinion of agent $i$'s friends to describe the ideological composition of their ego network:

$$x_{-i} = \frac{\sum_{j=1}^{n} a_{ij} x_j}{\sum_{j=1}^{n} a_{ij}} \qquad (2)$$

Quantity (2) does not provide comprehensive insight into agent $i$'s ideological environment, as the same values of $x_{-i}$ may—from the perspective of the baseline micro-assumptions—correspond to qualitatively different situations and lead to incorrect inferences (see Figure 3 for more details). For this reason, we also calculate the standard deviation of friends' opinions:

$$\sigma_{-i} = \sqrt{\frac{\sum_{j=1}^{n} a_{ij} \left( x_j - x_{-i} \right)^2}{\sum_{j=1}^{n} a_{ij}}}$$

This quantity measures the *diversity* of friends' opinions.

[Figure 3 is near here]

Note that even accounting for the diversity of friends' opinions may be insufficient, as different individuals may exert different influences on their peers. In this study, we assume that all influences are equal. It is important to note that our approach assumes "many-to-one" communication between users—but they may very well be engaging in other communication types (including one that is common online: "one-to-many"; Mäs, 2019). Users' opinions are restricted to the interval $[0, 1]$, which may also affect the results. For example, due to geometrical constraints, individuals with strong positions *a priori* have a lower chance to become stronger in their views than those with moderate positions. As such, we should separate individuals with strictly different positions while analyzing how they change their views in response to friends' opinions.

To avoid noisy opinion shifts, we only consider shifts that exceed a predefined value of 0.05, which is widely used in testing statistical hypotheses:

**Definition 1.** Opinion shift $x_i(t_k) \to x_i(t_{k+1})$ is *remarkable* if its magnitude $|x_i(t_{k+1}) - x_i(t_k)|$ is greater than 0.05.

**Definition 2.** *Estimated probability of opinion change* (EPOC) is defined as the proportion of individuals who make a remarkable opinion movement.

We calculate EPOC (and other metrics) for two time steps: $t_1 \to t_2$ and $t_2 \to t_3$.

Of great importance are situations where users shift their opinions toward the nearest edge of the opinion space. To emphasize these users, we introduce the following definition:

**Definition 3.** Agent $i$ *becomes stronger* in their opinion or *radicalizes* between times $t_k$ and $t_{k+1}$ if they make a remarkable opinion shift $x_i(t_k) \to x_i(t_{k+1})$ toward the nearest edge of the opinion space.

Note that Definition 3 is only correct for those users whose opinions differ from 0.5. However, we find that there are no such users in the dataset (for all three waves). We disentangle opinion shifts toward and away from the average opinion of friends:

**Definition 4.** Remarkable opinion shift $x_i(t_k) \to x_i(t_{k+1})$ is said to be *positive* if it is directed toward the friends' average opinion: $\left(x_i(t_{k+1}) - x_i(t_k)\right) * \left(x_{-i}(t_k) - x_i(t_k)\right) > 0$. Remarkable opinion shift is said to be *negative* if it is directed away from the friends' average opinion: $\left(x_i(t_{k+1}) - x_i(t_k)\right) * \left(x_{-i}(t_k) - x_i(t_k)\right) < 0$.

Note that both positive and negative shifts can result in radicalization. Among positive opinion shifts, we distinguish between those with and without skips:

**Definition 5.** Positive opinion shift $x_i(t_k) \to x_i(t_{k+1})$ is said to be *non-skipping* if new opinion $x_i(t_{k+1})$ lies in the closed interval between $x_i(t_k)$ and $x_{-i}(t_k)$. Positive opinion movement is said to be *skipping* if it is not non-skipping.

Note that negative, skipping, and non-skipping shifts collectively cover all possible remarkable opinion shifts. As such, EPOC can be naturally deconstructed as the sum of three terms standing for the probabilities of non-skipping, skipping, and negative shifts:

$$EPOC = EPOC_+ + EPOC_- = EPOC_+^{skip} + EPOC_+^{nonskip} + EPOC_-$$

In our study, we frequently explore different dependencies unfolding in the opinion space (for example, we consider the function $EPOC_+ = EPOC_+(x_i, x_{-i})$, which represents the estimated chance of a positive shift as a function of $x_i$ and $x_{-i}$), so the following two definitions are useful in analyzing them.

***Definition 6.*** Let us consider ideological group $G$, where $G \in \{SL, L, M, C, SC\}$. Equality $x_i = G$ indicates that agent $i$ belongs to ideological group $G$. Inequality $x_i > G$ means that agent $i$'s opinion is located further to the right than $G$ in the opinion space (for example, inequality $x_i > M$ means that $x_i = C$ or $x_i = SC$).

***Definition 7.*** Let us consider movement $G_1 \to G_2$, where $G_1, G_2 \in \{SL, L, M, C, SC\}$ and $G_1 < G_2$ (the case $G_2 < G_1$ is elaborated analogously). We determine the *area of negative shifts* by $x_{-i} < G_1$ and the *area of positive shifts* by $x_{-i} > G_1$. Inequality $G_1 < x_{-i} < G_2$ defines the *area of skipping movement*; if $G_2 \le x_{-i}$, we get the *area of non-skipping movements*. If $G_1 = G_2$, the shift $G_1 \to G_2$ is called *static*.

It is worthwhile to mention that static movements may constitute remarkable shifts. However, note that Definition 7 prohibits highlighting skipping movements if $G_1$ and $G_2$ are neighboring ideological groups.

### 5.2. Core Assumptions and Expectations

In Table 4, we summarize competing micro-assumptions that we test against real data in the following four subsections. For each micro-assumption, we detail how the answers to questions **Q1**, **Q2'**, and **Q3** should look like if the assumption were true. Note that we concentrate on how the data should look if we consider positive and negative shifts separately; we disentangle positive and negative shifts and analyze their properties in isolation. However, it is also important to understand how positive and negative shifts coexist in opinion dynamics. The literature suggests that positive influence is active when the opinion difference $|x_i - x_{-i}|$ is not too large; high $|x_i - x_{-i}|$ values likely result in negative influence. In other words, when opinion divergence is high enough, positive influence is replaced by negative influence. It means that high values of $|x_i - x_{-i}|$ should be associated with negative movements, whereas if $|x_i - x_{-i}|$ is not too large, then one should expect positive movements.

[Table 4 is near here]

In Table 4, we do not focus on the very important issue of how so-called "echo chambers" affect opinion radicalization. Echo chambers are connected groups of individuals with similar views and no immediate access to challenging content (Jasny et al., 2015). Scholars assert that individuals in echo chambers are more prone to radicalization (Del Vicario et al., 2017). However, the opinion-formation mechanisms listed in Table 4 cannot model the situation in which individuals surrounded by those with the same opinions radicalize; in such a case, under any of our core assumptions, individuals' opinions will be stable. To obtain polarization stemming from communication between individuals with similar opinions, one could, for example, clarify how these individuals communicate by specifying *arguments* in the model (Mäs & Flache, 2013). For the purposes of this study, an argument is a piece of information with a particular leaning. Once adopted by an individual, it plays a role in the development of their opinions. If two individuals with similar opinions provide each other with new arguments for the same position, they reinforce each other's opinions and, in turn, contribute to radicalization. If the theory that echo chambers reinforce radicalization is correct, we must expect individuals to radicalize more frequently when exposed to similar (or more radical but yet having the same bias) opinions.

### 5.3. Probability of Opinion Change

In this Subsection, we try to address the question "Under what conditions do individuals change their opinions more readily?" We start our analysis by calculating the map of cross-group movements (see Table 5). We find that the most popular user strategy is to stay within the current ideological group. The majority of cross-group shifts occur between neighboring ideological groups. Our analysis reveals that the populations of SLs and Ls grow in size, whereas SCs are nearly balanced between income and outcome; the populations of Ms and Cs explicitly go down.

[Table 5 is near here]

The next part of our analysis is structured around the investigation of EPOC as a function of both users' and friends' opinions: $EPOC = EPOC(x_i, x_{-i})$. For time step $t_1 \rightarrow t_2$, we observe 147,344 remarkable shifts (0.6 percent positive); for time step $t_2 \rightarrow t_3$, we observe 84,210 remarkable movements (0.579 percent positive). In Figure 4, we depict how EPOC varies with friends' average opinion $x_{-i}$ across different values of $x_i$, separated by movement type (positive or negative). To avoid situations in which both the focal user's opinion and the opinions of their friends are altered (since we do not know which change came first), we calculate EPOC by only considering individuals whose friends' average opinion changes by less than 0.05. In what follows, we calculate metrics only for such users. In Figure B1 (Appendix), we demonstrate how $EPOC(t_k \rightarrow t_{k+1})$ is moderated by the value of $|x_{-i}(t_{k+1}) - x_{-i}(t_k)|$ (the general trend is that higher values of $|x_{-i}(t_{k+1}) - x_{-i}(t_k)|$ increase the chance of opinion shift).

[Figure 4 is near here]

**Positive movements.** The probability of a positive movement is positively connected with the value of $|x_i - x_{-i}|$: if we bring $x_{-i}$ away from $x_i$ (left or right), $EPOC_+$ increases, excepting a number of zones near $x_{-i} = 0$ and $x_{-i} = 1$. The minima are observed at $x_{-i} \approx x_i$. In dependencies from Figure 4, one can highlight clear linear segments that are sometimes combined with sharp decreases at the edges of the opinion space.

**Negative movements.** The general trend is largely the same as that for positive movements: if we bring $x_{-i}$ away from $x_i$ (left or right), $EPOC_-$ increases, sometimes falling near the interval boundaries. One strict difference is that the minima are not located at $x_{-i} \approx x_i$; instead, they are usually slightly shifted from $x_i$. Some parts of curves have clear linear or inverted U-shaped forms.

**Mutual positioning of curves.** If $x_{-i} < x_i$, the probability of a positive shift is higher than that of a negative shift:

$$EPOC_+(x_i, x_{-i}) > EPOC_-(x_i, x_{-i})$$

If $x_{-i} > x_i$, the situation is rather unclear: for SLs and Ls, $EPOC_+$ usually dominates $EPOC_-(x_i, x_{-i})$; for other groups, we observe a minimal advantage in $EPOC_-$.

We also notice the following inequalities (see Figure B2 in Appendix):

$$EPOC_+(x_i = SL, x_{-i} = L) < EPOC_+(x_i = L, x_{-i} = SL);$$
$$EPOC_+(x_i = SL, x_{-i} = M) < EPOC_+(x_i = M, x_{-i} = SL);$$
$$EPOC_+(x_i = L, x_{-i} = M) < EPOC_+(x_i = M, x_{-i} = L);$$
$$EPOC_+(x_i = C, x_{-i} = M) > EPOC_+(x_i = M, x_{-i} = C);$$
$$EPOC_+(x_i = M, x_{-i} = SC) > EPOC_+(x_i = SC, x_{-i} = M);$$
$$EPOC_+(x_i = C, x_{-i} = SC) > EPOC_+(x_i = SC, x_{-i} = C).$$

(3)

Note that inequality $EPOC_+(x_i = L, x_{-i} = M) < EPOC_+(x_i = M, x_{-i} = L)$ in equation (3) holds only for time step $t_2 \rightarrow t_3$ (others are true for both time steps).

**Influence of opinion diversity.** We have not found any clear trends in how the diversity of friends' opinions affects the chance of opinion shift (see online supplementary materials).

**Skipping and non-skipping movements.** In an attempt to disentangle the nature of skipping (just 7 percent of all positive movements) and non-skipping movements, we first investigate how the value of $\sigma_{-i}$ differs between those who made skipping movements and those who made non-skipping movements. The intuition behind this is that skips may be attributed to high values of opinion diversity among friends (see Figure 3, panel c). However, we find a negligible difference (in both cases, the average of $\sigma_{-i}$ is approximately 0.141). We then plot the probabilities of making each of 25 possible (remarkable) movements in the opinion space (accounting for static movements but not imposing the requirement to be remarkable on them) as functions of $x_{-i}$, aiming to investigate how the character of a curve behaves while transitioning from the area of skipping movements to the area of nonskipping movements. If skipping movements are somewhat inadvertent, we should expect a sound change in behavior (presumably, growth with an increasing rate). Unfortunately, the dependencies (presented in Figure B3 in the Appendix and in the online supplementary materials) do not enable us to assert that skipping movements have a strictly different nature from non-skipping movements.

*5.4. Radicalization of Opinions*

The dependencies presented in Figure B3 enable us to investigate how the process of opinion radicalization is connected with users' opinions and their neighborhood ideological composition. To do so, we concentrate on two transitions that definitively meet the definition of radicalization (Definition 3): L -> SL and C -> SC. For convenience, we reproduce them separately in Figure 5. On the one hand, we do not consider individuals with initially strong positions, as they have little to no room in the opinion space for radicalization. On the other hand, we do not care about the radicalization of moderate individuals, as radicalization in these cases may be confused with political mobilization. Instead, we focus on identifying how environments can prompt individuals with clear political identities to become stronger in their views.

[Figure 5 in near here]

The curves have approximately the same character. For each, the probability of radicalization has two areas of relatively high values: (1) if $x_{-i} \in \{SL, L\}$ (for Ls) or $x_{-i} \in \{C, SC\}$ (for Cs) and (2) if $x_{-i} \in \{C, SC\}$ (for Ls) or $x_{-i} = SL$ (for Cs). Minima are reached at $x_{-i} \approx 0.5$. The radicalization of Ls is featured with the more pronounced minimum. It means that individuals radicalize more frequently (1) if they are exposed to similar views (that is, if they are in echo chambers) or more radical opinions yet having similar bias or (2) when they observe clearly opposite opinions. Users are less prone to radicalization if they are exposed to moderate views. We also observe that conservatives are less prone to radicalization than liberals.

**Influence of opinion diversity.** After controlling for opinion diversity (see Figure B4 in Appendix), we find that liberals embedded in networks with higher diversity (but a liberal or moderate average opinion) radicalize more frequently.

### *5.5. Magnitude of Opinion Change*

In this subsection, we analyze how the magnitude of opinion change is connected with users' and friends' opinions. Figure 6 illustrates how the magnitude of a remarkable shift depends on $x_{-i}$ after controlling for $x_i$ and movement type.

[Figure 6 is near here]

**Positive shifts.** Figure 6 reveals that the magnitude of positive opinion shifts tends to be positively correlated with the value of $|x_i - x_{-i}|$: friends who are more distant in the opinion space from the focal user tend to induce larger opinion shifts toward their positions. Observing dependencies, we emphasize segments with an inverted U-shaped form.

**Negative shifts.** We report that, for all users except Ms, the global maxima are reached at the nearest edges of the opinion space. If we bring $x_{-i}$ away from $x_i$ toward the furthest edge, the magnitude tends to change through an inverted U-shaped form.

**Influence of opinion diversity.** We obtain no clear picture of how opinion diversity is connected with opinion shift magnitude (see online supplementary materials).

### *5.6. Positive or Negative?*

The final part of our analysis concentrates on the following question: "Given we know that an individual will make a (remarkable) movement, will it be positive or negative?" This question has the following motivation. Assume that positive and negative shifts largely stem from interplay between positive influence and negative influence. How can we predict which component will operate for a given user (if they will be activated)? From a theoretical perspective, as we previously mentioned in Subsection 5.2, one should expect that if the distance between users' and friends' opinions is not too huge, a positive component is in charge. When $|x_i - x_{-i}|$ becomes large enough, positive influence gives way to negative influence, meaning we can expect a negative shift to be more likely.

We plot the ratio of the number of positive movements to the number of negative movements as a function of $x_{-i}$ across different values of $x_i$ (see Figure 7). It is clear that, if one increases the value of $|x_i - x_{-i}|$, the positive/negative ratio persistently features an inverted U-shaped form, indicating that positive shifts have a (relatively) higher chance to occur if the distance $|x_i - x_{-i}|$ is not too high and not too small.

[Figure 7 is near here]

## 6. Discussion

Two immediate conclusions from our findings are that users' opinions are generally stretched to the edges of the opinion space and the number of moderate users declines over time. In a nutshell, we observe opinion polarization. Note that the term "polarization" here refers only to the sustainable growth of two clusters of polar opinions ($x = 0$ and $x = 1$)—our setup restricts opinions to the interval $[0,1]$. We also observe a clear leftward trend in the opinion space. The drastic difference in user activity (measured as the number of remarkable movements) between time steps $t_1 \rightarrow t_2$ and $t_2 \rightarrow t_3$ may be explained by the political context: the presidential election was conducted between $t_1$ and $t_2$.

The way how the probability and average magnitude of a positive shift depend on users' and friends' opinions (precisely, on the distance between them) largely align with the linear and moderated (positive) assumptions, a result that is in line with existing empirical studies (see Table 1). Our findings on how the chance of a positive shift varies by users' and friends' opinion strength (inequalities in equation (3)) also match the theory and can be understood as follows: more radical individuals are less sensitive to peer influence than those with weaker positions. This explains (partially, at least—we should not forget about negative shifts and, of course, selectivity) both the structure of homophily (Table 3) and the map of cross-group opinion shifts (Table 5): individuals with strong opinions frequently attract their peers, inducing cascades of positive shifts. As a result, these individuals tend to accumulate an enormous nearby concentration of individuals with similar views. Note that dynamics of connections between users (that are beyond the scope pf this article) can also explain such patterns of homophily. For example, radical individuals are highly engaged in political discourse. As such, two radical users that follow similar information sources (and thus having similar estimated opinions) have a high chance to set a friendship connection through mutual conversations.

Note that the inequalities in equation (3) are quite non-trivial from the perspective of geometrical intuition. Let us consider, for example, the following inequality:

$$EPOC_+(x_i = SL, x_{-i} = L) < EPOC_+(x_i = L, x_{-i} = SL).$$

This means that the probability that a strong liberal user will move right (given their friends are, on average, liberals) is less than the probability that a liberal user will move left (given their friends are, on average, strong liberals). What is non-trivial here is the fact that, by default, SLs have far more space to move right than Ls do to move left. As a result, one might expect the former probability to be greater.

It is worthwhile to note that the role of assimilative influence in positive shifts is not entirely clear, as other effects may also contribute to positive shifts.

**Example 1.** The presence of a common stimulus (Aral & Nicolaides, 2017) is a potential confounder that may be confused with positive influence. Befriended users tend to follow similar information sources and, in turn, be influenced in a similar way; this effect is not accounted for in this study, as we assume that information sources serve only as indicators of individuals' opinions—not the sources. As a result, we may observe a positive shift stemming from the non-simultaneous reaction (simultaneous reaction is controlled for by only considering individuals whose friends' average opinion shifts by less than 0.05) of befriended individuals to the same stimuli.

Other examples of potential confounders that could contribute to the observed positive shifts are selectivity, shifts in the political bias of information sources, and the assumption of static ties (see Kozitsin, 2020).

However, alongside positive shifts, we also observe negative shifts that constitute a significant fraction of all remarkable movements. Our analysis reveals that negative shifts have several interesting properties. Figure 4 serves as clear evidence that SLs and SCs are associated with enormous rates of negative shifts among their peers. On the one hand, this observation is in line with the theory—the furthest opinions are the most repulsive. However, the theory suggests that negative influences should have a minimal likelihood if friends' opinions are close to those of the focal user. In contrast, we observe minima at $x_{-i}$ that differ from—but are close to—$x_i$. For all individuals except Ms, this may be attributed to geometrical constraints. For example, Cs apriori have more space to move left than to move right. As such, the lowest rate of negative movements for such individuals should be observed at $x_{-i} < C$. For Cs, the minima are observed at $x_{-i} = M$, the nearest leftward position. For Ms, we observe a clear divergence between theory and empirics that cannot be explained by geometrical constraints (Ms are located strictly in the middle; both directions should have equal priority). Nonetheless, magnitudes of negative opinion shifts tend to follow the theory (the moderated assumption). Global maxima in Figure 6 at $x_{-i}$, observed at the nearest edges (which violates the theory), may be explained by geometrical intuition: the magnitude of a negative shift likely has the highest value when an individual directs it toward the furthest edge.

The way how the chance and magnitude of a negative shift are connected to opinion divergence is somewhat non-trivial (in a way that, importantly, largely matches the theory); it may be considered as an argument in favor of repulsive influence. If repulsive influence is really the cause of negative shifts, Ms' behavior clearly indicates a relatively high rate of repulsive influence induced by individuals with a conservative bias. This observation may be attributed to political affairs (in the spirit of ideas presented by Böttcher & Gersbach (2020)). For instance, it may constitute a display of citizens' negative attitudes toward Russian pension reform (and, correspondingly, those who support the government that launched this reform) in 2018,[1] when this data was gathered.

More arguments in favor of the negative influence hypothesis are that (1) the negative influence is theorized to be more likely when observing attitude—rather than belief—dynamics (exactly our case) and (2) "there are psychological theories that predict that individuals develop increasingly positive views on a political candidate when they learn that members of the opposite social category dislike that person" (Mäs, 2019). One critical point against the negative influence hypothesis is that "negative influence may be very unlikely when individuals do not communicate face-to-face as students do in a classroom, but in a computer-mediated setting like a comment board on the Internet" (Mäs, 2019).

We must note, however, that observed patterns of negative shifts may be partially explained in a different way, without referring to the notion of repulsive influence.

---

[1] https://www.bbc.com/news/world-europe-45342721.

*Example 2 (explanation of patterns of negative shifts without negative influence).* Here, we provide an example of how to explain an enormous number of negative shifts among individuals in friendships with users holding strong positions. Let us assume that radical positions are associated with some degree of discomfort, that some users holding them are inclined to tone down their opinions. This discomfort could be driven by various elements of political affairs, such as political talk shows (Petrov & Proncheva, 2020). As a result, individuals with strong positions are pushed to reconsider their views. Additionally, let us suppose that estimations of users' opinions are not fully correct (this assumption is actually quite realistic). Befriended individuals tend to have similar real views. As a result, we should expect the following regularity: the more friends that agent $i$ has with a particular position $G \in \{SL, L, M, C, SC\}$, the higher the chance that agent $i$ has the same real position (regardless of agent $i$'s estimated position). If $G = SL$ or $G = SC$, there is a higher chance of individuals toning down their real opinion; in turn, this may lead to subscriptions to information sources that shift agent $i$'s estimated opinion away from their friends' (estimated) opinions

The next question is one of the balance between positive and negative shifts. Our results indicate that, for a given individual with opinion $x_i$, left-located opinion $x_{-i} < x_i$ is more likely to be associated with a positive shift than with a negative shift. Interestingly, if a (remarkable) movement has occurred for an individual with opinion $x_i$, that movement is highly likely to be positive if the value of opinion divergence $|x_i - x_{-i}|$ is not too small and not too high. The prevalence of negative shifts for high $|x_i - x_{-i}|$ values is consistent with the theory; their prevalence if $x_i$ and $x_{-i}$ are close may be attributed, for example, to users striving for uniqueness (Mäs et al., 2010). In fact, this result gives another perspective on those of Kozitsin (2020), who found that, after controlling for user opinion, the probability of shifting left varies with the average opinion of the user's friends $x_{-i}$, and the target function has a near-sinusoidal wave-shaped form whereby if source opinion $x_{-i}$ is less than user opinion $x_i$, the probability is greater than that for $x_{-i} > x_i$.

Despite failing to disentangle skipping and non-skipping movements, we put forward the following potential explanation for skipping movements based on the profound nature of the considered process.

*Example 3 (explanation of skipping movements).* Let us consider agent $i$ making shift $M \rightarrow SC$ in the opinion space. This shift is essentially driven by some change in agent $i$'s subscriptions, presumably as a result of influence from a friend, source $j$. Let us assume that source $j$ follows public pages $P_1, P_2$ and $P_3$: the first two ($P_1$ and $P_2$) with a conservative bias and the third ($P_3$) with a strong conservative bias. Naturally, source $j$'s estimated opinion should be rather conservative. If agent $i$ starts to follow only $P_3$ (perhaps influenced by source $j$'s repost from $P_3$), their opinion may be estimated as strongly conservative. As a result, we observe a skip in the opinion space performed by agent $i$ over source $j$'s opinion.

Finally, we have found mixed support for the theory that echo chambers reinforce radicalization. More specifically, we have demonstrated that individuals have a high chance to radicalize if they (1) are surrounded by individuals with similar or stronger views yet having the same bias or (2) are exposed to opposing radical views. After observing any kind of radical opinion—liberal or conservative—individuals tend to radicalize with more enthusiasm. The proliferation of OSNs may facilitate the formation of echo chambers, as OSNs are driven by algorithms that are supposed to accommodate communication between like-minded individuals (Bakshy et al., 2015; Geschke et al., 2019; Kozyreva et al., 2020; Mäs & Bischofberger, 2015; Perra & Rocha, 2019; Rossi et al., 2019). Scholars continue to assess whether personalization systems increase opinion radicalization (that, in turns, may lead to opinion polarization): if echo chambers do really drive radicalization, then personalization systems likely contribute to opinion radicalization. Our results only partially confirm these ideas. We argue that the most efficient way to mitigate radicalization is to avoid exposing individuals to radical opinions, instead exposing them only to moderate positions.

## 7. Conclusion

This paper presents an empirical study of the opinion dynamics of a large-scale sample of online social network users. Users' opinions were estimated via their subscriptions to information sources, and we have analyzed how friendship connections affect the dynamics of these estimations.

The majority of existing empirical studies suggest that the formation of continuous opinions (presented as continuous quantities) largely follow the linear positive mechanism whereby individuals tend to modify their opinions toward those of their friends with a magnitude proportional to the pre-existing difference in opinions (Friedkin et al., 2021; Takács et al., 2016). However, some papers suggest a moderated positive influence: there is a maximum rate of influence if the difference is not too small and not too high; otherwise, it decreases (Moussaïd et al., 2013). Studies that reported a negative influence (opinion shifts away from the opinions of peers) either have methodological concerns (Knippenberg et al., 1990; Mazen & Leventhal, 1972) or are based on natural experiments in which researchers are unable to control for all possible confounding factors (Liu & Srivastava, 2015).

In this paper, we distinguish between positive (toward friends' opinions) and negative (away from friends' opinions) opinion shifts. We find that the properties of positive shifts are largely in line with the predictions of existing theoretical models of opinion formation. Precisely, the existence and magnitude of positive opinion shifts are positively related (largely through a linear form or an inverted U-shaped form) to the degree of divergence in opinions between user and their friends. Interestingly, we find that both the chance and magnitude of negative shifts increase alongside the opinion divergence between the focal user and their friends (largely through, again, a linear form or an inverted U-shaped form). The balance between positive and negative shifts is also moderated by opinion divergence: if the opinions of the focal user and their friends are too similar or dissimilar, there is a relatively low chance of a positive shift.

These results challenge existing concerns about the presence of a negative influence; they constitute a clue for the context in which researchers should search for such an influence. However, we also demonstrate how to explain our empirical results without referring to the notions of positive and negative influence. Concurrently, we investigate how the process of radicalization is connected to the ideological heterogeneity of users' neighborhoods. We argue that individuals are less likely to radicalize if they are exposed to moderate positions; in contrast, radical views—regardless of their bias—induce a relatively high degree of radicalization.

One crucial aspect that was not identified in this research is how users create and delete ties with each other. The evolution of the social graph has a potentially significant impact on opinion dynamics as its structure determines largely how information flows between users. A greater understanding as to how individuals form connections at the micro-level can improve our knowledge of social graph formation processes (M. E. J. Newman, 2001; Snijders, 2017). Additionally, it would be interesting to control for users' demographic characteristics, such as age, gender, and location, as they may affect how influence processes unfold on OSNs (Kovanen et al., 2013). For example, individuals with similar characteristics likely have more communication contacts; additionally, age and gender may control for sensitivity to peer influence (Peshkovskaya et al., 2019).

The findings presented in this research shed some additional light on the problem of linking theoretical models of opinion formation with real data. However, our findings should not be overestimated due to a large number of dataset limitations. From this perspective, it would be interesting to evaluate the extent to which these limitations may affect the outcome that is visible to an observer (who analyzes the data). One could introduce a model in which agents form connections with each other and subscribe to information sources. These sources would have their own political biases and agents would decide whether to follow a particular information account by comparing this bias with their own opinion and considering how many of their friends follow this account (in the fashion of complex contagion models (D. Centola, 2010; Damon Centola, 2019; Christakis & Fowler, 2013; Mønsted et al., 2017; Ugander et al., 2012)). Of course, it would be beneficial to allow agents to follow information sources whose biases differ from agents' views. In this model, an observer would estimate agents' opinions using their subscriptions to information sources with some error (also a parameter that could be varied) and then analyze these estimations, as we have done in this paper. By considering the different rules as to how agents communicate with each other (including the impact of personalization algorithms) and how they form connections (with other agents and information sources), we can obtain different outcomes and compare them against the real data.

**Acknowledgments**

The authors are grateful to anonymous reviewers for their invaluable comments.

## Appendix

## Appendix A. Testing Interrelation Between Ideological Position, Age, and Number of Friends

We fix two public pages in VK that are official media accounts: "Медуза" (Meduza)[2] and "Новости RT на русском" (RT)[3]. RT is a state-controlled Russian media outlet that is often accused of being the "Kremlin's propaganda tool"; many argue that "Russia spreads disinformation via RT."[4] In contrast, Meduza is an online newspaper that is not controlled by the state; in fact, according to state-controlled outlets, Meduza generally opposes the current Russian Government.[5] If the estimations of users' opinions are correct, we should expect that an individual following only RT is estimated as one with conservative polarity $x_{RT} > 0.5$ while an individual following only Meduza is estimated as one with liberal polarity $x_M < 0.5$. To check this, we create two artificial users that meet the above requirements and estimate their opinions. We find that the user following Meduza has $x_M = 0.38$ while the one following RT has $x_{RT} = 0.56$; these results confirm our expectations.

Next, we download a new, independent snapshot of Meduza and RT followers. We consider only active users (at least one platform interaction per month) who are Russians of at least 18 years of age with open privacy settings. We collect information on 117,736 users following only RT, 84,283 users following only Meduza, and 3,656 users who follow both accounts. We observe that Meduza followers are, on average, younger and have more friends than RT followers (see Figure A1). However, we find a non-trivial dependency between user age and number of friends within follower groups. For those who follow only RT, there is a weak negative correlation (Pearson correlation coefficient equal to -0.078, p-value near zero). For those who follow only Meduza, there is a weak positive correlation (0.096, p-value near zero). For those who follow both Meduza and RT, there is an insignificant (p-value equal to 0.09) and negligible negative correlation (-0.02). This suggests that our initial hypothesis holds only for conservative individuals; liberal users with more friends are generally older than those with fewer friends. This observation may be explained as follows. From Figure A1, we can see that the audience of Meduza is dominated by persons under 40. Recalling that VK was established in 2006, we can hypothesize that it should have the most popularity among those who were young in the mid-2000s. However, young people today may divert their attention to newer platforms (e.g., TikTok, Clubhouse); thus, they may not be relatively active on VK (but still more active than people over 40, who constitute a remarkable portion of RT followers).

---

[2] https://meduza.io/en

[3] https://www.rt.com/

[4] https://archives.cjr.org/feature/what_is_russia_today.php, https://www.cbsnews.com/news/russian-news-english-accent-11-12-2005/, https://www.theguardian.com/commentisfree/2019/jul/26/russia-disinformation-rt-nuanced-online-ofcom-fine

[5] https://www.rt.com/op-ed/377715-rt-gazillionaire-media-funding/, https://www.rt.com/russia/443562-meduza-editor-sexual-harassment/

[Figure A1 is near here]

**Appendix B. Additional Figures and Tables**

[Figure B1 is near here]

[Table B1 is near here]

[Table B2 is near here]

[Table B3 is near here]

[Figure B2 is near here]

[Figure B3 is near here]

[Figure B4 is near here]

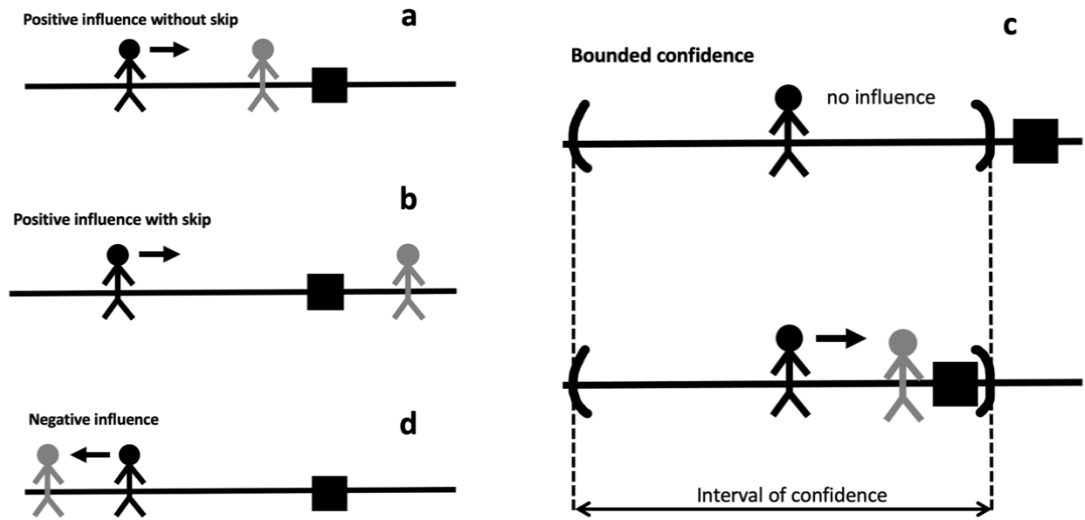

*Figure 1.* In this figure, we concentrate on how an agent shifts their opinion on one-dimensional opinion space—a horizontal axis—under different influence assumptions. The black icons represent agents' past opinions; the grey icons represent their new positions formed in response to an influence (black square).

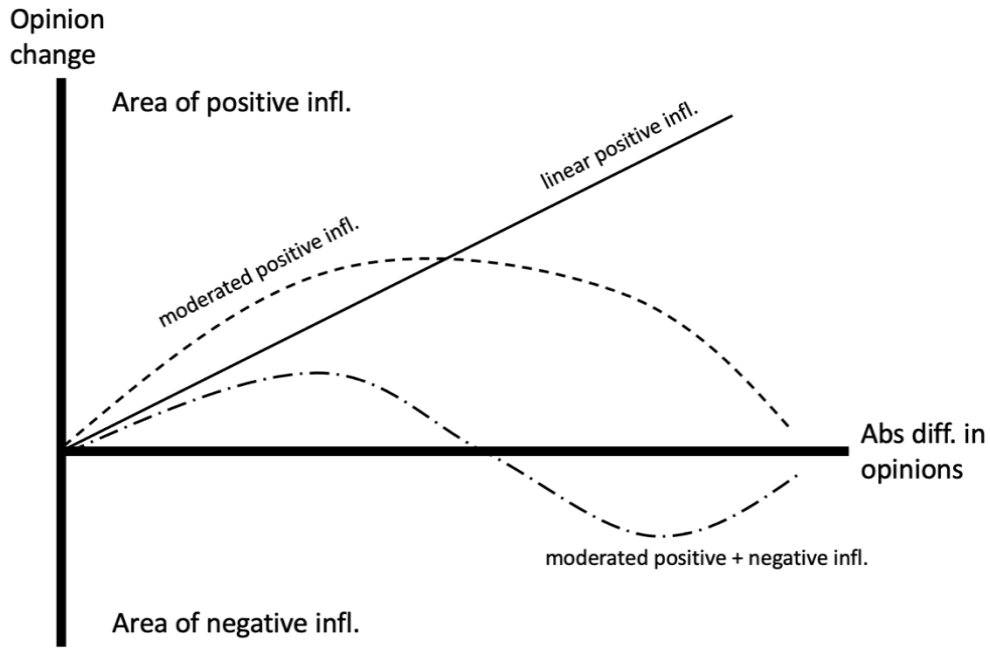

*Figure 2.* Here, we plot how the magnitude of opinion change $x_i(t+1) - x_i(t)$ may vary by the absolute difference between an individual's opinion $x_i(t)$ and that of the influence source $x_j(t)$. The upper half-plane represents opinion shifts toward the influence source $((x_i(t+1) - x_i(t)) * (x_j(t) - x_i(t)) > 0)$; the lower half-plane represents negative influence cases in which opinion shifts away from the influence source $x_j(t)$.

Table 1

Studies testing influence assumptions against real data (listed in order of publication)

| Reference | Approved or Rejected Hypothesis | | |
|---|---|---|---|
| | Linear positive | Moderated positive | Pos.+Neg. |
| (Mazen & Leventhal, 1972) | | | +* |
| (Knippenberg et al., 1990) | | | +** |
| (Krizan & Baron, 2007) | | | - |
| (Moussaïd et al., 2013) | | + | |
| (Liu & Srivastava, 2015) | | | +*** |
| (Kerckhove et al., 2016) | + | | |
| (Takács et al., 2016) | + | | |
| (Clemm von Hohenberg et al., 2017) | + | | |
| (Friedkin & Bullo, 2017) | + | | |
| (Friedkin et al., 2021) | + | | |

Note: * - methodological concern: did not to consider trends in opinion shifts; ** - methodological concern: did not disentangle positive in-group influence and negative out-group influence; *** - nonlinear in both cases with an increasing rate.

Table 2

Information on ideological groups (at time moment $t_2$)

| Group | SLs | Ls | Ms | Cs | SCs |
|---|---|---|---|---|---|
| Opinion interval | [0, 0.2) | [0.2, 0.4) | [0.4, 0.6) | [0.6, 0.8) | [0.8, 1] |
| Population | 125,714 | 312,063 | 874,327 | 266,509 | 70,216 |
| Avg number of friends | 25.68 | 21.2 | 17.13 | 13.03 | 10.43 |

Table 3

Average ideological composition of neighborhoods across different ideological groups (at time moment $t_2$).

| Ideological group | Avg fraction of users in neighborhood | | | | |
|---|---|---|---|---|---|
| | SLs | Ls | Ms | Cs | SCs |
| SLs | 0.15 | 0.24 | 0.46 | 0.11 | 0.03 |
| Ls | 0.11 | 0.23 | 0.51 | 0.12 | 0.03 |
| Ms | 0.08 | 0.20 | 0.54 | 0.15 | 0.03 |
| Cs | 0.07 | 0.17 | 0.49 | 0.21 | 0.06 |
| SCs | 0.07 | 0.15 | 0.45 | 0.24 | 0.10 |
| Null model | 0.08 | 0.19 | 0.53 | 0.16 | 0.04 |

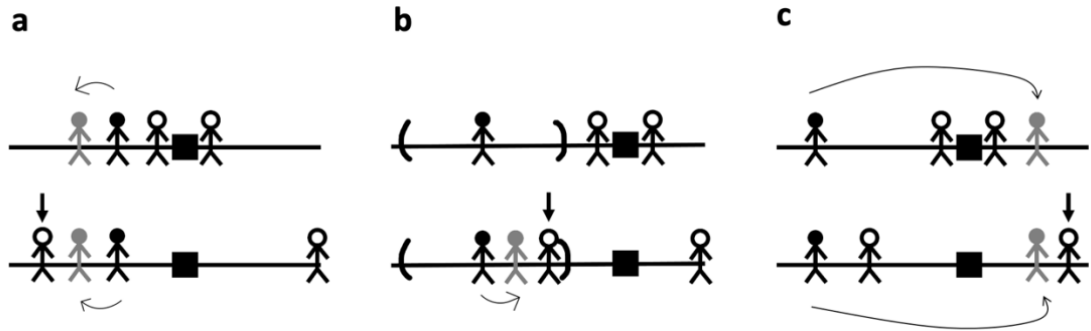

*Figure 3.* Illustration of how opinion diversity among peers may affect individuals'
opinion dynamics. The black icons represent agents' past opinions; the grey icons
represent their new positions formed in response to an influence (black square). The icons
with non-colored heads represent friends' opinions. **Panel a.** In the top case, the agent
shifts away from their friends' opinions (potentially as a result of negative influence). In
the bottom case, the agent shifts toward the arrowed opinion—the real source of (positive)
influence. In both cases, observers (those who do not know about the agent's friends and
see only their average opinion) likely conclude the existence of a negative influence.
**Panel b.** Assume that opinion dynamics obey the bounded confidence assumption. In the
top case, there is no chance for influence, as both of the potential influencers are beyond
the agent's interval of confidence. In the bottom case, one of the influencers falls within
the interval of confidence and exerts a positive influence. The initial settings in both cases
are the same for observers—but the outcomes are different. **Panel c.** In the top case, the
focal agent makes a skip over both of the potential influencers. In the bottom case, one of
the influencers (the one located further away from the agent) is a real source of positive
influence (without a skip). In both cases, observers witness a skip.

Table 4

Competing micro-assumptions and expectations

| Micro-assumption | | EPOC | Radicalization | Magnitude |
|---|---|---|---|---|
| Positive influence | Linear | Grows linearly with the value of $\|x_i - x_{-i}\|$ | Likely when $0.5 < x_i < x_{-i}$ or $x_{-i} < x_i < 0.5$ | Grows linearly with the value of $\|x_i - x_{-i}\|$ |
| | Moderated | First increases then decreases as $\|x_i - x_{-i}\|$ increases (inverted U-shaped form) | Likely when $0.5 < x_i < x_{-i}$ or when $x_{-i} < x_i < 0.5$ and $\|x_i - x_{-i}\|$ has a "moderate" value | First increases then decreases as $\|x_i - x_{-i}\|$ increases (inverted U-shaped form) |
| Negative influence | Linear | Grows linearly with the value of $\|x_i - x_{-i}\|$ | Likely when $x_{-i} < x_i$ and $0.5 < x_i$ or when $x_{-i} > x_i$ and $0.5 > x_i$ | Grows linearly with the value of $\|x_i - x_{-i}\|$ |
| | Moderated | First increases then decreases as $\|x_i - x_{-i}\|$ increases (inverted U-shaped form) | Likely when $x_{-i} < x_i$ and $0.5 < x_i$ or when $x_{-i} > x_i$, $0.5 > x_i$, and $\|x_i - x_{-i}\|$ has a "moderate" value | First increases then decreases as $\|x_i - x_{-i}\|$ increases (inverted U-shaped form) |

Note 1: The word "moderate" is somewhat fairly vague here, meaning only that $\|x_i - x_{-i}\|$ is not too small and not too large (depending on context).

Note 2: Magnitudes of negative movements unlikely follow the linear positive assumption due to geometrical constraints: for high values of $\|x_i - x_{-i}\|$ there is little to no room to make a negative shift. This deduction is not applicable for the chance of a negative shift, excepting the cases of ideological groups located near the opinion space edges (as they have no space to radicalize).

Table 5

Map of movements.

| | | -> SLs | -> Ls | -> Ms | -> Cs | -> SCs | Income | Outcome |
|---|---|---|---|---|---|---|---|---|
| $t_1 \rightarrow t_2$ | SLs | 113,081 | 5,662 | 991 | 117 | 25 | 12,633 | 6,795 |
| | Ls | 11,141 | 269,961 | 21,032 | 448 | 49 | 42,102 | 32,670 |
| | Ms | 1,307 | 35,724 | 830,633 | 16,706 | 233 | 43,694 | 53,970 |
| | Cs | 147 | 636 | 21,245 | 244,116 | 4,863 | 22,393 | 26,891 |
| | SCs | 38 | 80 | 426 | 5122 | 65046 | 5,170 | 5,666 |
| $t_2 \rightarrow t_3$ | SLs | 121,036 | 3,915 | 645 | 98 | 20 | 8,865 | 4,678 |
| | Ls | 8,071 | 291,304 | 12,483 | 186 | 19 | 30,433 | 20,759 |
| | Ms | 690 | 26,107 | 835,230 | 12194 | 106 | 27,909 | 39,097 |
| | Cs | 83 | 351 | 14,454 | 247,695 | 3,926 | 15,983 | 18,814 |
| | SCs | 21 | 60 | 327 | 3,505 | 66,303 | 4,071 | 3,913 |

Note: each cell represents the population of users who make a particular movement.

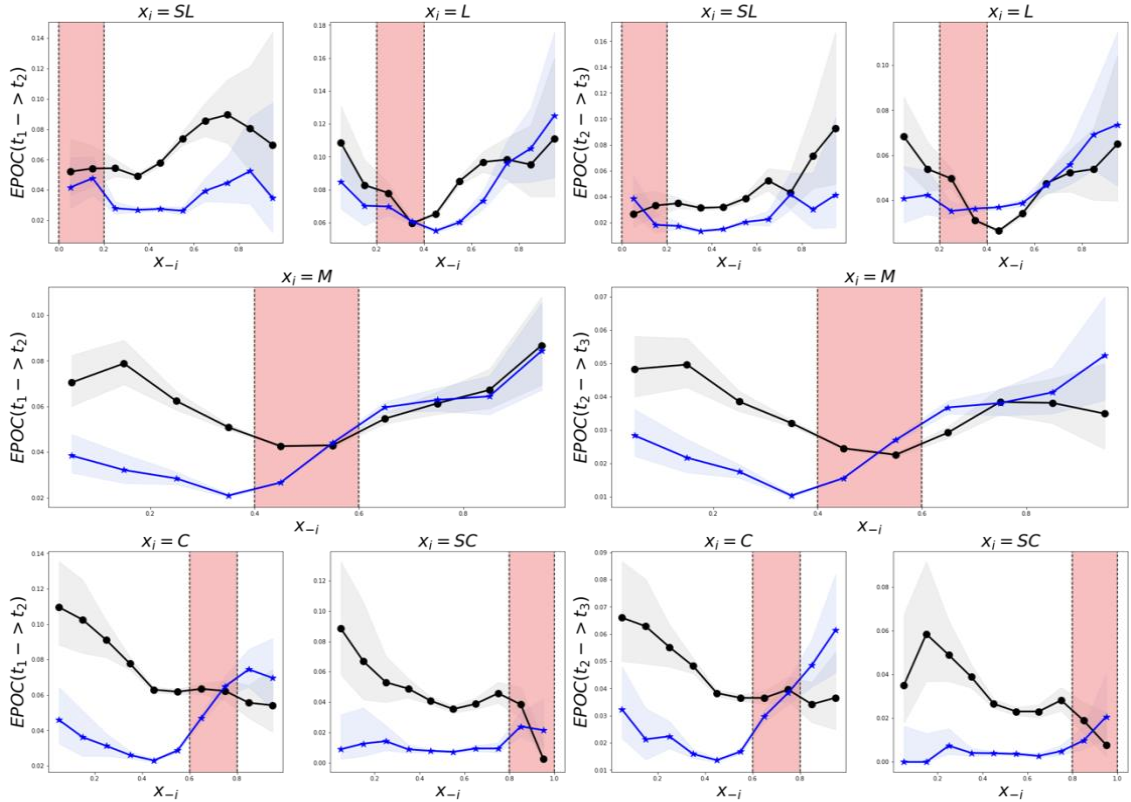

*Figure 4.* EPOC as a function of friends' average opinion $x_{-i}$ across ideological groups, separated by movement type. Colored columns represent the value of $x_i$. Black dots plot $EPOC_+$; blue stars plot $EPOC_-$. **Positive movements.** $EPOC_+$ is positively connected with the distance $|x_i - x_{-i}|$. The minima are located at $x_{-i} \approx x_i$. We highlight clear linear segments (for example: $t_1 \to t_2, x_i = M, x_{-i} \in [0.6,1]$; $t_1 \to t_2, x_i = SC, x_{-i} \in [0,0.4]$; $t_2 \to t_3, x_i = C, x_{-i} \in [0,0.4]$) and sharp decreases at the edges of the opinion space (for example: $t_1 \to t_2, x_i = SL, x_{-i} \in [0.7,1]$; $t_2 \to t_3, x_i = SC, x_{-i} \in [0,0.2]$). For non-moderate individuals we notice intervals where $EPOC_+$ does not change considerably ($t_1 \to t_2, x_i = SL, x_{-i} \in [0.2,0.4]$; $t_1 \to t_2, x_i = C, x_{-i} \in [0.4,0.6]$; $t_2 \to t_3, x_i = SC, x_{-i} \in [0.4,0.8]$). **Negative movements.** If we bring $x_{-i}$ away from $x_i$ (left or right), $EPOC_-$ tends to increase, sometimes falling near the interval boundaries. Minima are not located at $x_{-i} \approx x_i$; instead, they are usually slightly shifted from $x_i$ (for SLs and Ls – rightward, for Ms, Cs and SCs - leftward). Some parts of curves have clear linear (for example: $t_1 \to t_2, x_i = L, x_{-i} \in [0.5,1]$; $t_1 \to t_2, x_i = C, x_{-i} \in [0,0.5]$; $t_2 \to t_3, x_i =$

$M, x_{-i} \in [0,0.35]$) or inverted U-shaped ($t_1 \rightarrow t_2, x_i = SL, x_{-i} \in [0.7,1]$; $t_2 \rightarrow t_3, x_i = SC, x_{-i} \in [0,0.4]$) forms. The latter can be explained by geometrical constraints (see Table 4, Note 2). **Mutual positioning of curves.** We see that if $x_{-i} < x_i$, $EPOC_+$ is higher than $EPOC_-$.

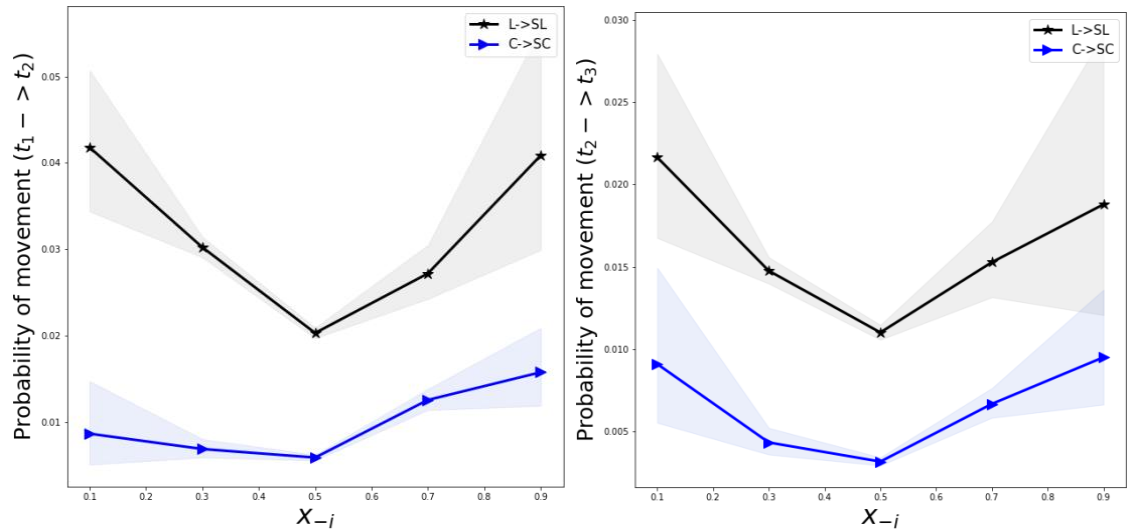

*Figure 5.* Probabilities of movements L -> SL and C -> SC as functions of $x_{-i}$. The left panel represents time step $t_1 \rightarrow t_2$; the right panel represents time step $t_2 \rightarrow t_3$. We highlight two areas of relatively high values of the radicalization rate: (1) if $x_{-i} \in \{SL, L\}$ (for Ls) or $x_{-i} \in \{C, SC\}$ (for Cs) and (2) if $x_{-i} \in \{C, SC\}$ (for Ls) or $x_{-i} = SL$ (for Cs). Minima are reached at $x_{-i} \approx 0.5$. The radicalization of Ls is featured with the more pronounced minimum. Figure serves as a clear evidence that conservatives are less prone to radicalization than liberals.

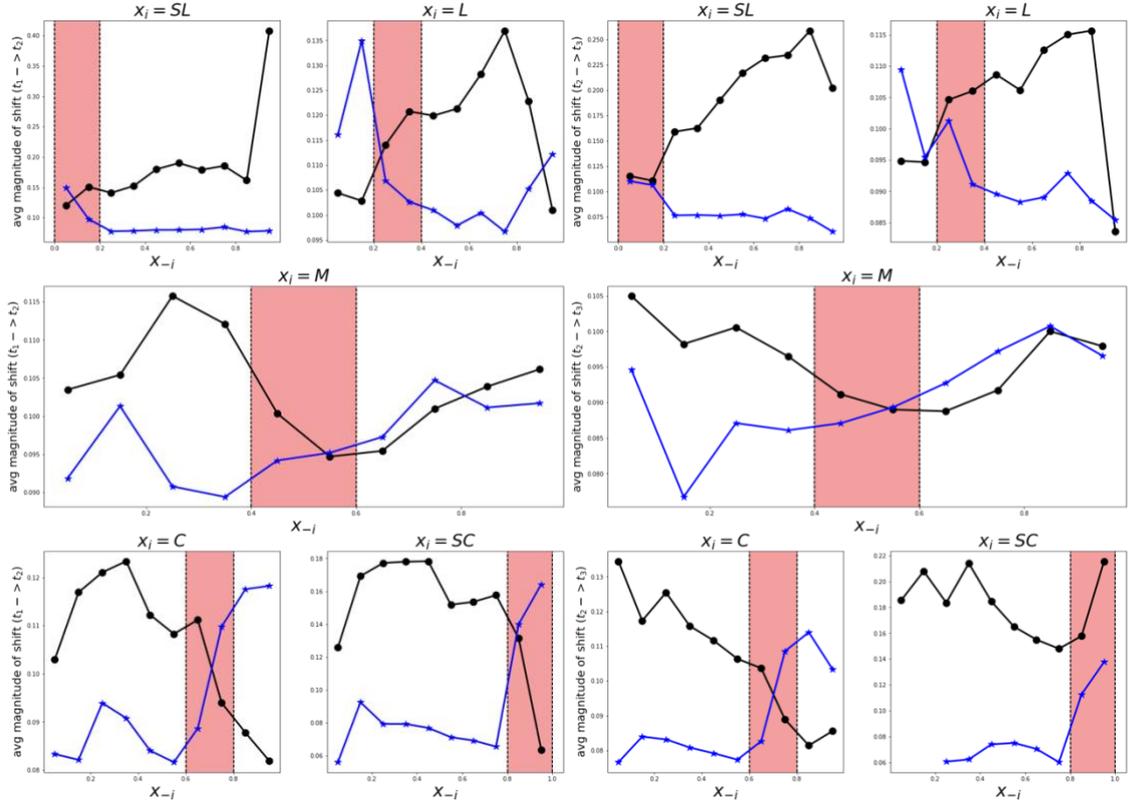

*Figure 6.* Average opinion shift magnitude as a function of $x_{-i}$ across different values of $x_i$, separated by movement type. We consider only remarkable shifts. Left panels represent time step $t_1 \rightarrow t_2$; right panels represent time step $t_2 \rightarrow t_3$. Colored areas represent the value of $x_i$. Black dots plot magnitudes of positive movements; blue starts plot magnitudes of negative movements. **Positive shifts.** The magnitude of positive opinion shifts tends to be positively correlated with the value of $|x_i - x_{-i}|$, largely through an inverted U-shaped form. **Negative shifts.** For all users except Ms, the global maxima are reached at the nearest edges of the opinion space. If we bring $x_{-i}$ away from $x_i$ toward the furthest edge, the magnitude tends to change through an inverted U-shaped form.

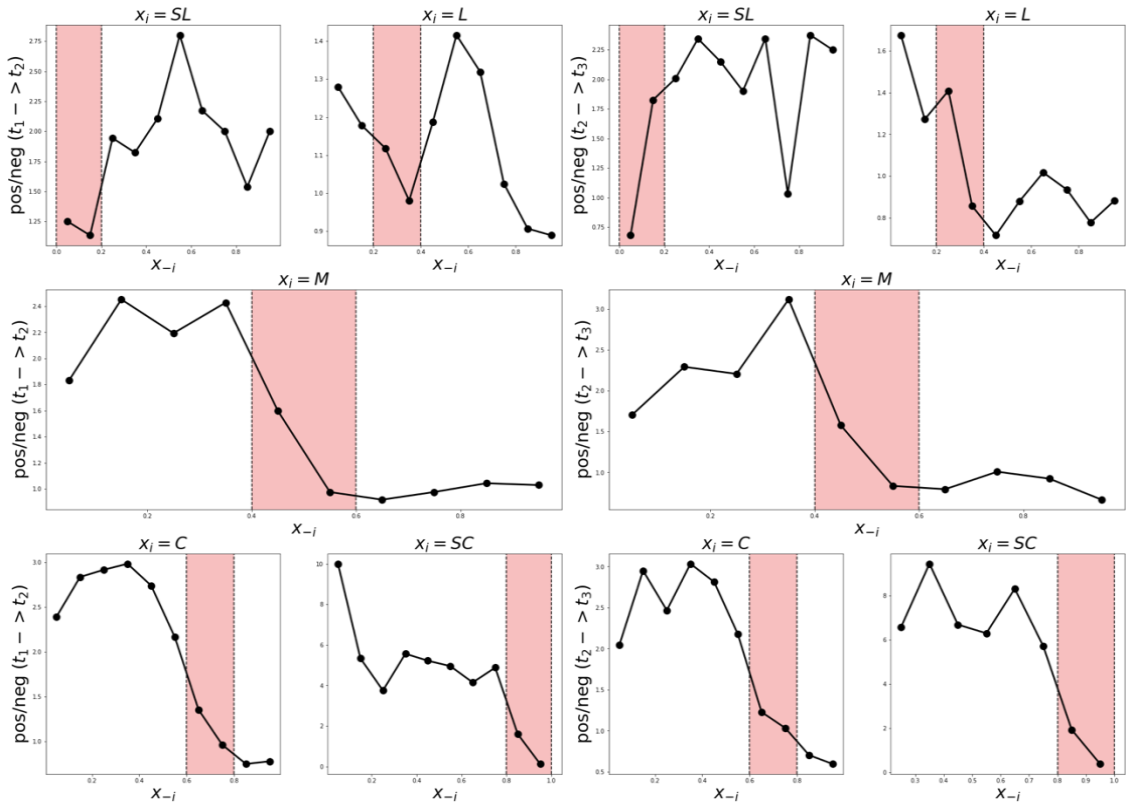

*Figure 7.* Positive/negative ratio as a function of $x_{-i}$, separated by $x_i$. Left panels represent time step $t_1 \rightarrow t_2$; right panels represent time step $t_2 \rightarrow t_3$. Colored areas represent the value of $x_i$. If one increases the value of $|x_i - x_{-i}|$, the positive/negative ratio persistently features an inverted U-shaped form, indicating that positive shifts have a (relatively) higher chance to occur if the distance $|x_i - x_{-i}|$ is not too high and not too small.

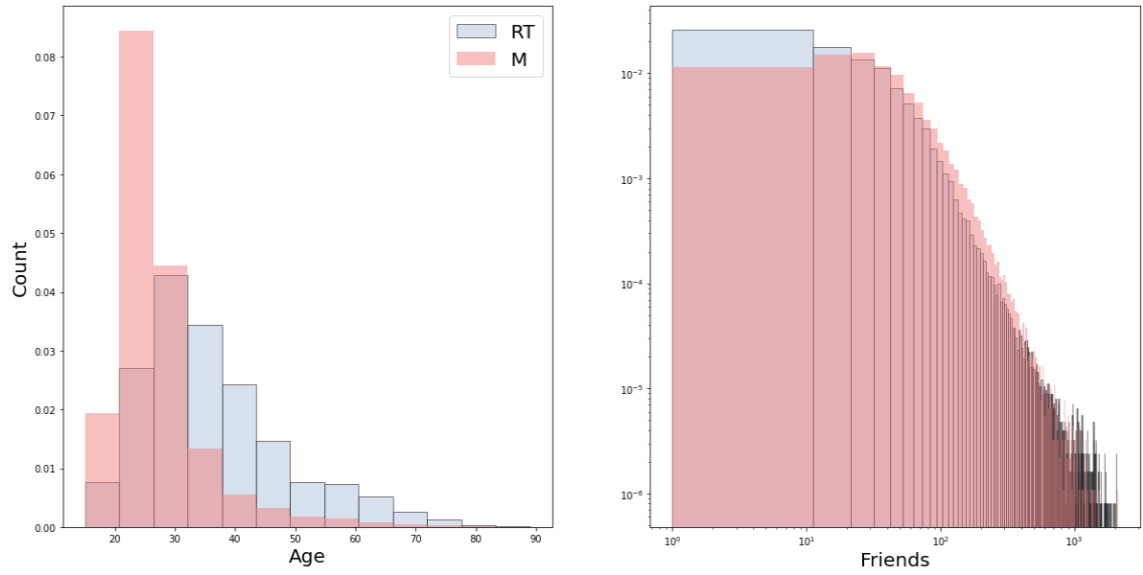

*Figure A1.* Density histograms representing the age structure (left panel) and friend distribution (right panel) of Meduza and RT followers.

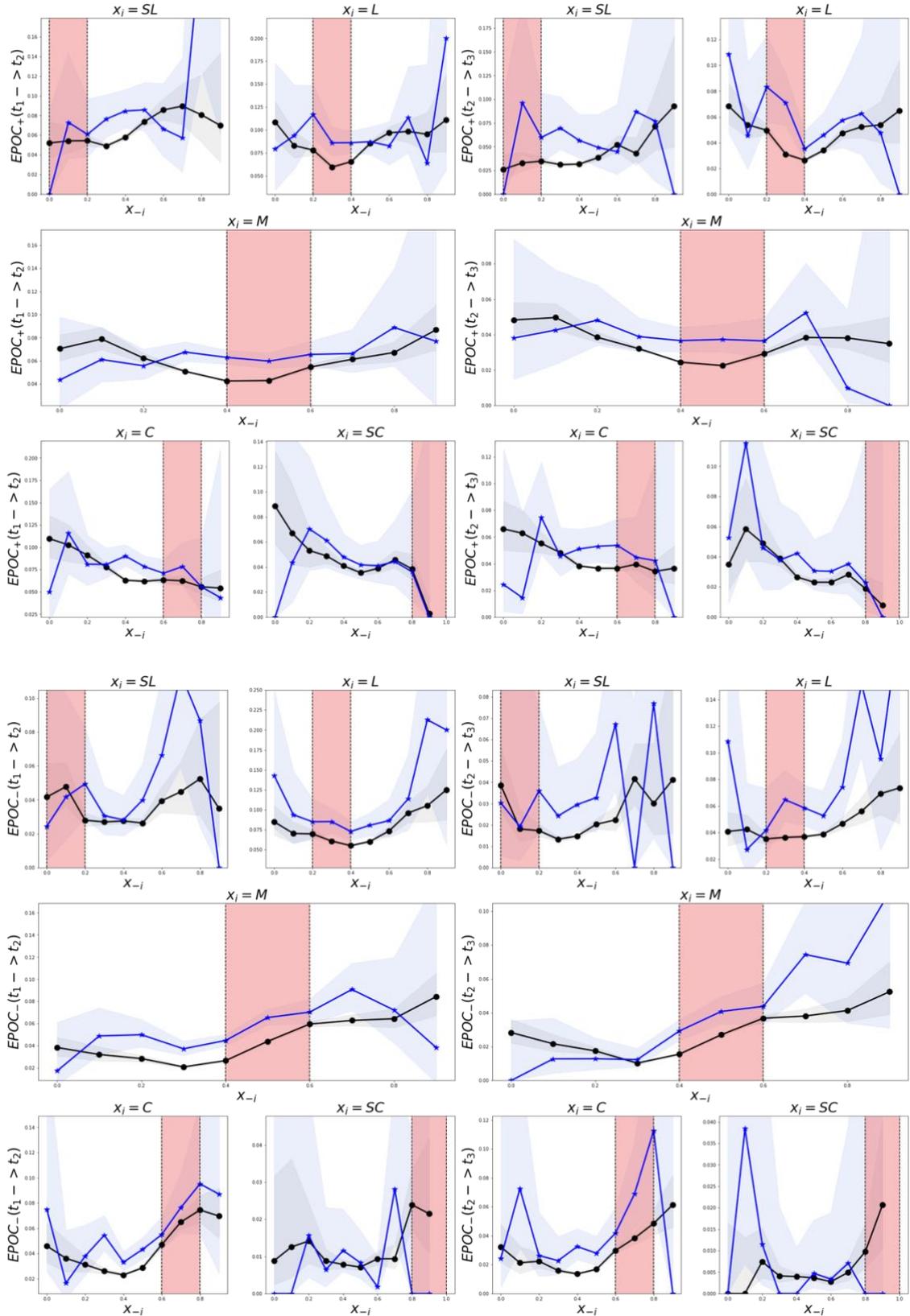

*Figure B1.* In this plot, we demonstrate how $EPOC(t_k \to t_{k+1})$ is moderated by the value of $|x_{-i}(t_{k+1}) - x_{-i}(t_k)|$. Black dots plot EPOC if $|x_{-i}(t_{k+1}) - x_{-i}(t_k)|$ is under 0.05;

blue starts plot EPOC if $|x_{-i}(t_{k+1}) - x_{-i}(t_k)| \geq 0.05$. We see that higher values of $|x_{-i}(t_{k+1}) - x_{-i}(t_k)|$ are correlated with a higher chance of an opinion shift.

Table B1

Average ideological composition of neighborhoods across different ideological groups among users with fewer than five friends (at time moment $t_2$).

| Ideological group | Avg fraction of users in neighborhood | | | | |
|---|---|---|---|---|---|
| | SL | L | M | C | SC |
| SL | 0.14 | 0.22 | 0.44 | 0.15 | 0.05 |
| L | 0.09 | 0.21 | 0.51 | 0.15 | 0.04 |
| M | 0.07 | 0.18 | 0.53 | 0.18 | 0.05 |
| C | 0.06 | 0.15 | 0.46 | 0.24 | 0.09 |
| SC | 0.06 | 0.13 | 0.40 | 0.27 | 0.14 |
| Null model | 0.08 | 0.19 | 0.53 | 0.16 | 0.04 |

Table B2

Average ideological composition of neighborhoods across different ideological groups among users with fewer than 25 friends and more than or equal to five friends (at time moment $t_2$).

| Ideological group | Avg fraction of users in neighborhood | | | | |
|---|---|---|---|---|---|
| | SL | L | M | C | SC |
| SL | 0.15 | 0.24 | 0.47 | 0.12 | 0.03 |
| L | 0.10 | 0.22 | 0.52 | 0.13 | 0.03 |
| M | 0.08 | 0.20 | 0.54 | 0.15 | 0.03 |
| C | 0.07 | 0.17 | 0.50 | 0.20 | 0.05 |
| SC | 0.07 | 0.16 | 0.47 | 0.23 | 0.08 |
| Null model | 0.08 | 0.19 | 0.53 | 0.16 | 0.04 |

Table B3

Average ideological composition of neighborhoods across different ideological groups among users with more than or equal to 25 friends (at time moment $t_2$).

| Ideological group | Avg fraction of users in neighborhood | | | | |
|---|---|---|---|---|---|
| | SL | L | M | C | SC |
| SL | 0.17 | 0.27 | 0.47 | 0.08 | 0.02 |
| L | 0.13 | 0.26 | 0.50 | 0.09 | 0.02 |
| M | 0.11 | 0.23 | 0.54 | 0.11 | 0.02 |
| C | 0.09 | 0.20 | 0.53 | 0.15 | 0.03 |
| SC | 0.08 | 0.20 | 0.51 | 0.17 | 0.04 |
| Null model | 0.08 | 0.19 | 0.53 | 0.16 | 0.04 |

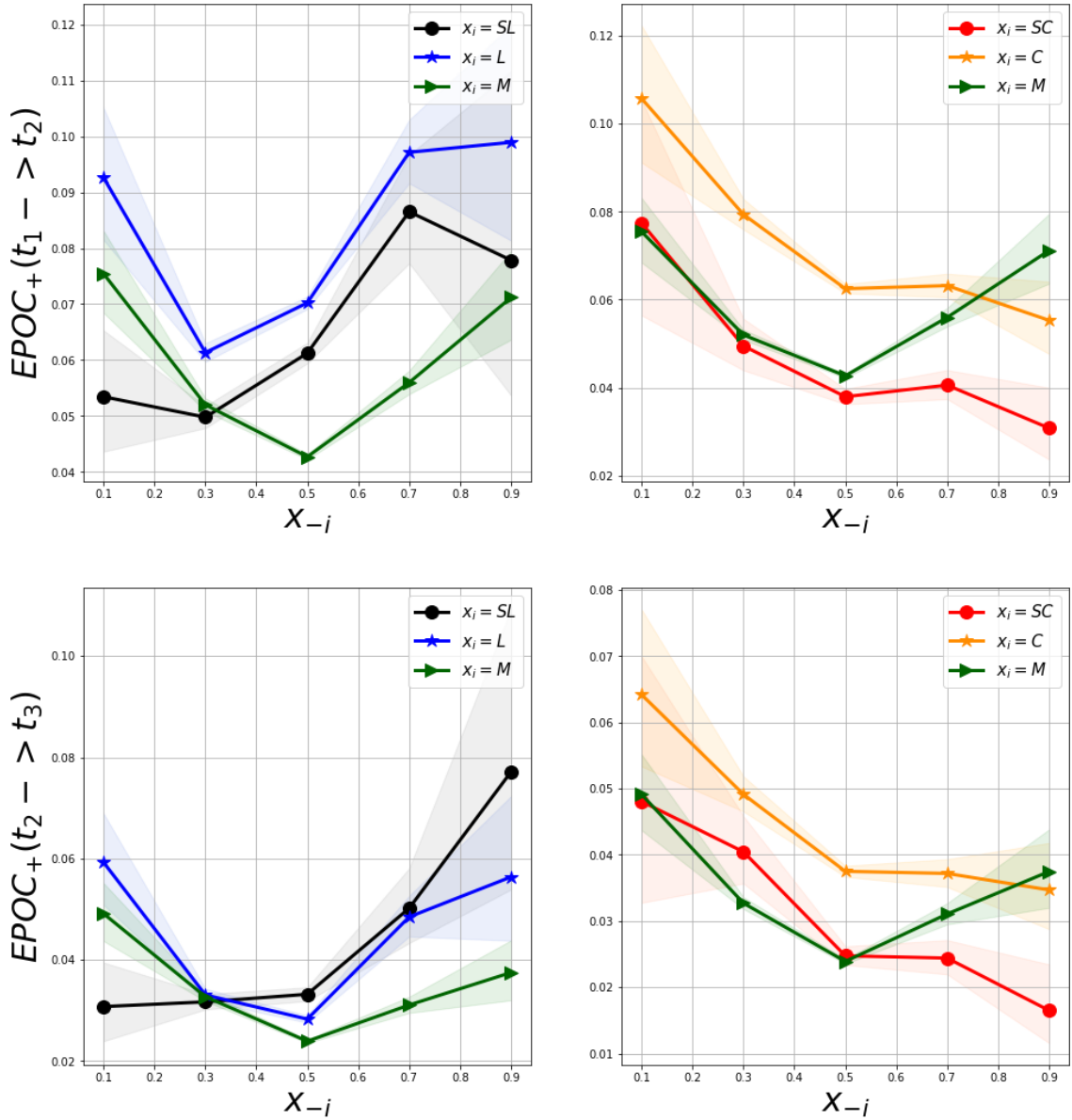

*Figure B2.* $EPOC_+$ as a function of $x_{-i}$, separated by $x_i$. We emphasize the following

regularity. For two groups $G_1$ and $G_2$, where without loss of generality $G_1 < G_2$, it tends

to hold that $EPOC_+(x_i = G_1, x_{-i} = G_2) < EPOC_+(x_i = G_2, x_{-i} = G_1)$ if $G_2 \leq M$ and

$EPOC_+(x_i = G_1, x_{-i} = G_2) > EPOC_+(x_i = G_2, x_{-i} = G_1)$ if $G_1 \geq M$. Exceptions here

are: (1) Ls and Ms for time step $t_1 \to t_2$; Ms and Cs for both time steps.

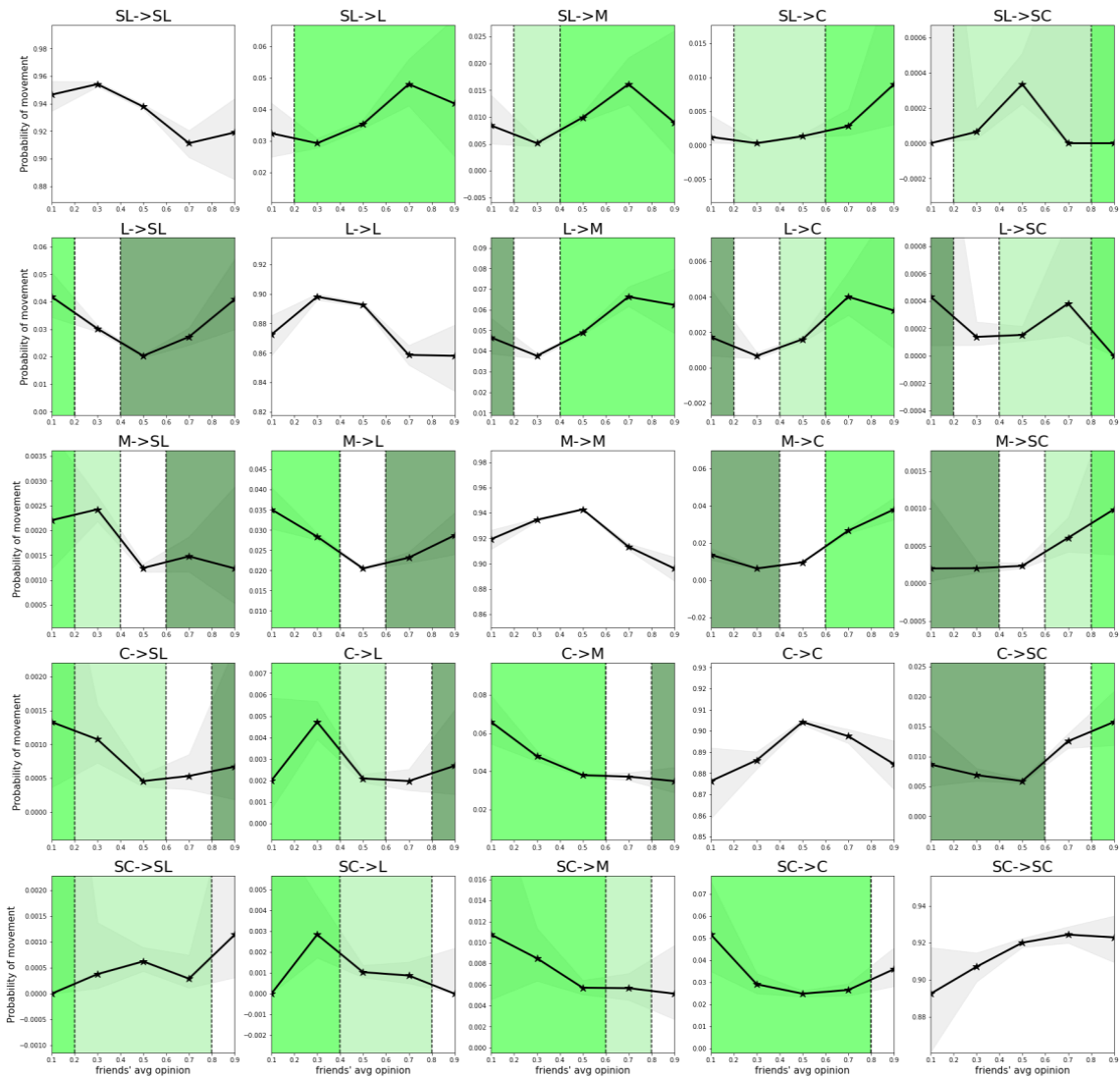

*Figure B3.* These subplots represent how the probability of a particular opinion shift depends on the average opinion of friends (here, we consider time step $t_1 \rightarrow t_2$; the same subplots for time step $t_2 \rightarrow t_3$ are included in the online supplementary materials). The dashed lines illustrate the boundaries between areas of negative (dark green), positive skipping (light green) and positive nonskipping (lime green) movements. Some subplots demonstrate that the curves do change their behavior after transitioning from the area of skipping movements to the area of nonskipping movements (see, for example, SL -> C or SC -> M whereby the curves start to grow with a higher rate after entering the nonskipping zone). However, the similar behavior (growth with an increasing rate) may be observed for movements that imply no skips (C -> M, SC -> C).

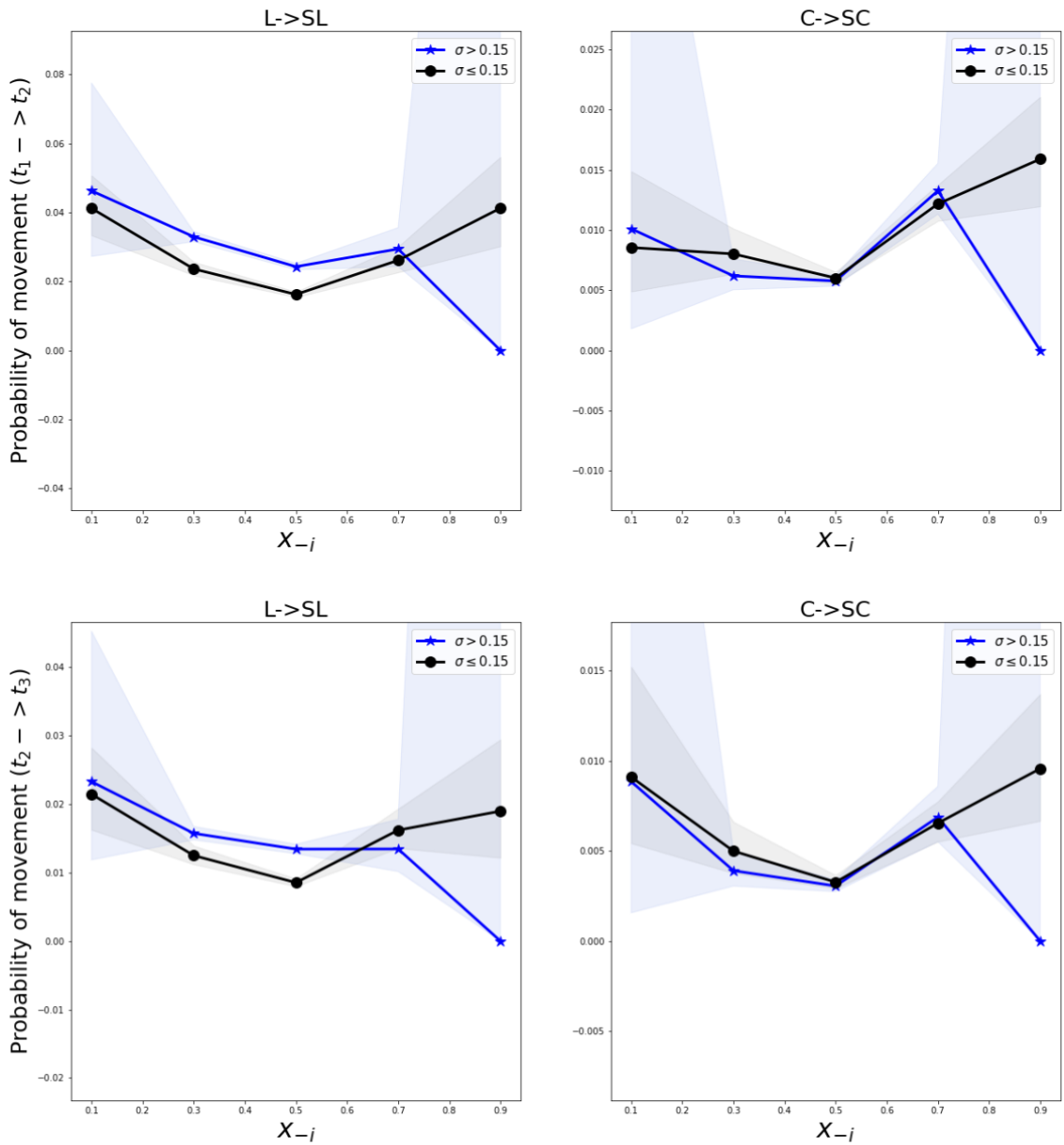

*Figure B4.* Probability of opinion shifts L -> SL and C -> SC as functions of $x_{-i}$, separated by $\sigma_{-i}$. Liberals embedded in networks with higher diversity (but a liberal or moderate average opinion) radicalize more frequently.